\documentclass[12pt,preprint]{aastex}

\usepackage{amsmath}

\shorttitle{CO$_2$ formation in quiescent clouds}
\shortauthors{Noble et al.}

\begin{document}


\title{CO$_2$ formation in quiescent clouds; an experimental study of the CO + OH pathway.}


\author{J. A. Noble} 
\affil{Department of Physics, Scottish Universities Physics Alliance, University of Strathclyde, Glasgow, G4 ONG, Scotland.}
\affil{LERMA-LAMAp, Universit\'{e} de Cergy-Pontoise, Observatoire de Paris, ENS, UPMC, UMR 8112 du CNRS, 5 mail Gay Lussac, 95000 Cergy Pontoise Cedex, France.}

\author{F. Dulieu}
\affil{LERMA-LAMAp, Universit\'{e} de Cergy-Pontoise, Observatoire de Paris, ENS, UPMC, UMR 8112 du CNRS, 5 mail Gay Lussac, 95000 Cergy Pontoise Cedex, France.}
\email{francois.dulieu@obspm.fr}

\author{E. Congiu}
\affil{LERMA-LAMAp, Universit\'{e} de Cergy-Pontoise, Observatoire de Paris, ENS, UPMC, UMR 8112 du CNRS, 5 mail Gay Lussac, 95000 Cergy Pontoise Cedex, France.}

\and

\author{H. J. Fraser}
\affil{Department of Physics, Scottish Universities Physics Alliance, University of Strathclyde, Glasgow, G4 ONG, Scotland.}



\begin{abstract}
The formation of CO$_2$ in quiescent regions of molecular clouds is not yet fully understood, despite CO$_2$ having an abundance of around 10--34~\%~H$_2$O. We present a study of the formation of CO$_2$ via the non-energetic route CO + OH on non-porous H$_2$O and amorphous silicate surfaces. Our results are in the form of temperature-programmed desorption spectra of CO$_2$ produced via two experimental routes: O$_2$ + CO + H and O$_3$ + CO + H. The maximum yield of CO$_2$ is around 8~\% with respect to the starting quantity of CO, suggesting a barrier to CO + OH. The rate of reaction, based on modelling results, is 24 times slower than O$_2$ + H. Our model suggests that competition between CO$_2$ formation via CO + OH and other surface reactions of OH is a key factor in the low yields of CO$_2$ obtained experimentally, with relative reaction rates $k_{CO+H} \ll k_{CO+OH} < k_{H_2O_2+H} < k_{OH+H},k_{O_2+H}$. Astrophysically, the presence of CO$_2$ in low A$_V$ regions of molecular clouds could be explained by the reaction CO + OH occurring concurrently with the formation of H$_2$O via the route OH + H.
\end{abstract}



\section{Introduction}\label{sec-intro}

The first observations of solid CO$_2$ (henceforth CO$_{2(s)}$) were made by the InfraRed Astronomical Satellite (IRAS, \citet{dHendecourt89}). The molecule has since been observed in numerous environments, including towards galactic centre sources \citep{deG96}, massive protostars \citep{ger99,gib04}, low mass YSOs \citep{num01,pon08}, background stars \citep{kne05}, and in other galaxies \citep{shi10,oli11}. Based on these observations, CO$_{2(s)}$ is seen to be seemingly ubiquitous, and one of the most abundant solid phase molecular species, approximately 10--34~\%~H$_2$O. It is believed to form in the solid phase, due to low gas phase abundances \citep{vanD96}, with evidence suggesting that much CO$_{2(s)}$ production occurs in quiescent regions \citep{pon06, num01}, yet the key question remains: how does CO$_{2(s)}$ form?

Many experimental studies have been performed to study the energetic formation routes to CO$_2$. Irradiation of pure CO ices with photons \citep{ger96}, charged particles \citep{pal98}, and electrons \citep{jam06} have yielded CO$_2$. Similar experiments with mixtures of CO and H$_2$O were also successful \citep{ehr97, pal98, iop09, laf10}. The irradiation of hydrogenated carbon grains with ions and electrons produced small quantities of CO and CO$_2$ \citep{men04,men06}.

CO$_{2(s)}$ is abundant in quiescent, as well as star-forming regions. While the role of energetic pathways can not be discounted entirely in these regions \citep{whi98}, the study of non-energetic formation routes is fundamental to fully understanding the observed abundances of CO$_{2(s)}$. Potential non-energetic formation routes \citep{ruf01} are:
\begin{equation}\label{eqn:CO+O}
  CO + O \rightarrow CO_2
\end{equation}
\begin{equation}\label{eqn:CO+OH}
  CO + OH \rightarrow CO_2 + H
\end{equation}
\begin{equation}\label{eqn:HCO+O}
  HCO + O \rightarrow CO_2 + H
\end{equation}
\begin{equation}\label{eqn:HCO+OH}
  HCO + OH \rightarrow CO_2 + H_2
\end{equation}

Theoretical studies of route (\ref{eqn:CO+O}) suggest that the formation of CO$_2$ proceeds with a high barrier of around 2500 -- 3000~K, lowered on surfaces via the hot O atom or Eley-Rideal mechanisms \citep{tal06, gou08}. A solid phase study determined that this pathway was feasible only via reaction in water pores, under a water ice cap, and upon heating \citep{ros01}, suggesting that it would not occur under the conditions present in quiescent molecular clouds. Reactions (\ref{eqn:HCO+O}) and (\ref{eqn:HCO+OH}) have never been studied expressly in the solid phase. 

Route (\ref{eqn:CO+OH}) has been extensively studied in the gas phase, both experimentally \citep{fro93, ful96, bau05} and theoretically \citep{yu01, chen05, sun08}, due to its importance in atmospheric and combustion chemistry. Recently, reaction (\ref{eqn:CO+OH}) was experimentally studied in the solid phase by Reflection Absorption InfraRed Spectroscopy (RAIRS) for the first time, with positive results \citep{oba10}. Due to the limitations of the adopted method, it was not possible to produce a pure beam of OH, and therefore the chemistry is difficult to constrain with a simple series of reactions; in particular, it was experimentally complex to distinguish between (\ref{eqn:CO+OH})--(\ref{eqn:HCO+OH}). OH was produced in the gas phase via a plasma discharge of H$_2$O, a process which yields a mixture of products including OH, H, H$_2$, O and O$_2$. Although it is claimed that all OH radicals are in the rovibrational ground state due to collisions with the beam walls, well defined spectroscopic studies of plasma discharges suggest that interaction with the walls is likely to lead to OH recombination, rather than yield ground state OH, and that the major components of a plasma of H$_2$O are H$_2$ and H$_2$O, with lower abundances of OH \citep{med02, fuj02}. Furthermore, experiments in the absence of CO produced H$_2$O$_2$ and O$_3$, whose yields varied with surface temperature, suggesting that surface temperature itself, mobility of the discharge products on the surface, and, potentially, the desorption rate of CO from the surface, rather than the rovibrational state of the OH, is responsible for the changing yields of CO$_2$ observed at different temperatures. Finally, reaction (\ref{eqn:CO+OH}) was found to proceed with little or no barrier, suggesting the presence of rovibrationally excited OH. We contend that, under these conditions, the CO$_2$ yield can not be assumed to be independent of the excitation state of OH and thus further study of reaction (\ref{eqn:CO+OH}) is imperative. A subsequent RAIRS study produced OH in the solid phase from a mixture of O$_2$:CO in a multilayer regime \citep{iop10}. Due to the multilayer regimes investigated, both previous studies also involved more complex chemistry than simply CO$_2$ formation, such as the formation of H$_2$CO$_3$ and CH$_3$OH, which further complicate the quantitative analysis of reaction (\ref{eqn:CO+OH}).

Here we present the first temperature-programmed desorption (TPD) spectra of CO$_{2(s)}$ formed via (\ref{eqn:CO+OH}) in the solid phase, under interstellar conditions of temperature and pressure. In this study, OH was produced on the surface by reaction of O$_2$ and O$_3$ with H in order to constrain better the reaction pathways in the system. The reaction was studied on both an amorphous silicate and a non-porous water surface, in a low coverage regime, with the aim of limiting chemistry to only CO$_2$ production. In contrast to previous studies, a simple kinetic model was developed to determine relative reaction efficiencies and calculate the activation energy of reaction (\ref{eqn:CO+OH}).


\section{Experimental}\label{sec-expt}

Experiments were performed using the FORMOLISM apparatus \citep{Amiaud06}. Briefly, the experimental set-up consists of an ultra high vacuum (UHV) chamber (base pressure $\sim$~10$^{-10}$~mbar), containing an amorphous silicate-coated copper surface (5--400~K, \citet{lem10}). Molecules are dosed onto the surface via two triply differentially pumped beam lines. Desorption of molecules from the surface is monitored using a quadrupole mass spectrometer (QMS), positioned directly in front of the surface. Experiments were performed on either bare silicate, or a non-porous, amorphous water film (\emph{np}-H$_2$O) of $\sim$100~monolayers (ML) grown on the silicate by spraying water vapour from a microchannel array doser (held at 120~K during water desorption, then cooled to 10~K before commencing the experiments).

Two different surfaces were investigated in order to, firstly, mimic two interstellar environments and, secondly, to determine the surface dependency of route (\ref{eqn:CO+OH}). Amorphous silicate is an appropriate mimic of interstellar dust grains, composed of siliceous and carbonaceous material \citep{gre02}. In molecular clouds these grains are covered in an ice mantle, the largest component of which is H$_2$O, at abundances of up to 100~ML \citep{wil02}. \emph{np}-H$_2$O was used in this study to eliminate the complexity of chemistry occurring in pores.

Neither O$_3$ nor O$_2$ has been observed in an interstellar ice, so these experimental conditions are not directly astrophysically relevant, but were used to produce OH in a controlled, reproducible manner. It is experimentally complex to create, maintain, and deposit onto a surface, a pure, stable beam of the OH radical in the ground state. Thus, in this work, OH was produced on the surface via two routes: the hydrogenation of O$_2$ and of O$_3$. O$_2$ is easier to utilise experimentally, but a study involving both species constrains reaction mechanisms better than using a single species. Due to the limits of sensitivity of the QMS, quantities below 0.1~ML were not investigated. 

OH was not measured directly on the surface, due to its short lifetime, but the production of O$_2$, H$_2$O$_2$ and H$_2$O, during control experiments on the hydrogenation of O$_2$ and O$_3$, confirms its presence, according to the reaction scheme:

\begin{equation}\label{eqn:O3+H}
  O_3 + H \rightarrow O_2 + OH
\end{equation}
\begin{equation}\label{eqn:O2+XH}
O_2 \xrightarrow{H}{} HO_2 \xrightarrow{H}{} H_2O_2 \xrightarrow{H}{} H_2O + OH
\end{equation}

Figure~\ref{fig-network} describes the chemical network of OH relevant to these experiments. The OH radical, produced by (\ref{eqn:O3+H}) and (\ref{eqn:O2+XH}), could react with CO as in (\ref{eqn:CO+OH}) to produce CO$_2$, or with H to form H$_2$O, thus, the relative rate of these reactions is an important factor determining the yield of CO$_2$ and the aim of this work was to elucidate the relative rates of these reactions. 

All experiments are summarised in Table~\ref{tbl1}; approximately 1.5~ML of O$_3$ or 0.5~ML of O$_2$ were dosed onto the surface via one molecular beam, followed by $\sim$0.5~ML of isotopically labelled $^{13}$CO via a second beam. Finally, H atoms were deposited, via a plasma discharge of H$_2$ on beam 1, for a range of exposure times of between 0 and 20 minutes. The deposition rate of H on the surface was $\sim$~5~x~10$^{12}$ atoms cm$^{-2}$s$^{-1}$, taking into account the dosing pressure and the dissociation rate of H$_2$ \citep{ami07}. During O$_3$ deposition, the surface was held at 45~K to ensure that any traces of O$_2$ present in the O$_3$ beam desorbed from the surface (for detailed O$_3$ production method, see \citet{mok09}); it was then cooled to 10~K before continuing. For all other molecules, the surface was held at 10~K during dosing. 

H atoms were hot when produced in the plasma discharge of H$_2$, but cooled to room temperature before exiting the molecular beam, due to collisions with the walls. Isotopically labelled $^{13}$CO was used to avoid contamination of the results by $^{12}$CO$_2$ or $^{12}$CO, pollutants present at very low gaseous concentrations in the chamber. 

To measure the products of the reaction, the surface was heated from 10 to 100~K; desorbing molecular species were monitored with the QMS. Each TPD cycle lasted approximately three hours.


\section{Results and Discussion}\label{sec-results}

\begin{figure*}
\plotone{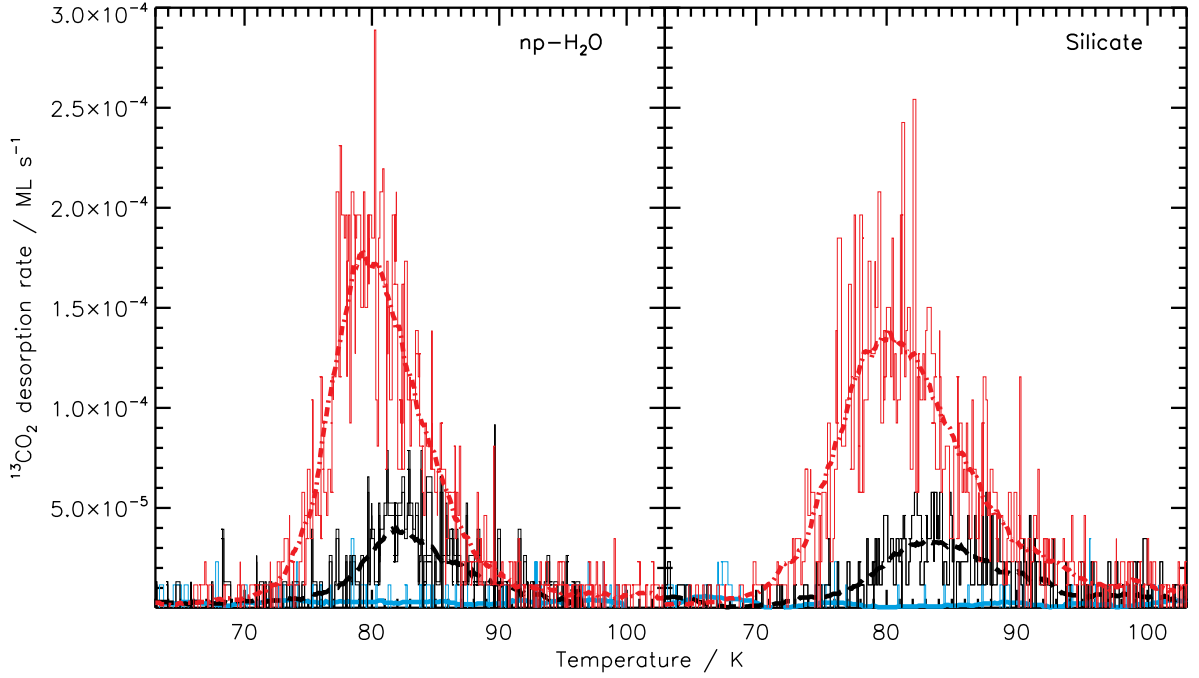}
\caption{Temperature programmed desorption spectra of Mass 45 ($^{13}CO_{2}$). The left panel shows desorption from a water surface while the right panel, a bare silicate surface. Experimental data are plotted in their raw form, accompanied by a smoothed version to guide the eye. Curves are labelled as follows: solid blue line, experiments A and H (0.5~ML O$_2$, $\sim$~0.5~ML $^{13}CO$, no H irradiation); dashed black line, experiments C and I (0.5~ML O$_2$, $\sim$~0.5~ML $^{13}CO$, 20 minutes H irradiation); dot-dashed red line, experiments G and K ($\sim$~1~ML O$_3$, $\sim$~0.5~ML $^{13}CO$, 20 minutes H irradiation). Production of $^{13}CO_{2}$ is seen for both starting molecules (O$_2$ and O$_3$) on both surfaces but O$_3$ yields significantly more $^{13}CO_{2}$ than O$_2$. There is no discernible surface dependence of the reaction under current experimental conditions.\label{fig-tpd}}
\end{figure*}

Figure~\ref{fig-tpd} shows TPD spectra of $^{13}$CO$_2$ produced by H irradiation of O$_3$ with $^{13}$CO, and O$_2$ with $^{13}$CO. $^{13}$CO$_2$ was produced during all experiments where H irradiation was performed. The $^{13}CO_{2}$ desorbs from each surface over a similar temperature range, but with a slightly different peak shape, indicative of the roughness of the underlying surfaces. It is clear that, since all data in Figure~\ref{fig-tpd} are measured at the same H irradiation time, substantially more $^{13}$CO$_2$ is produced by the reaction of O$_3$ than O$_2$. This is evident from reactions (\ref{eqn:O3+H}) and (\ref{eqn:O2+XH}): three hydrogenation steps are required to generate a single OH radical from O$_2$, whereas O$_3$ also generates an OH directly. 

As the solid blue line in Figure~\ref{fig-tpd} shows, if the experiment was conducted without H irradiation, no $^{13}$CO$_2$ was produced.  Nor was $^{13}$CO$_2$ production seen during control TPDs of $^{13}$CO, O$_2$ or O$_3$. When $^{13}$CO was not present, H$_2$O and H$_2$O$_2$ formed, rather than $^{13}$CO$_2$. It was assumed that all the $^{13}$CO$_2$ formed during H irradiation, in agreement with previous experiments \citep{oba10,iop10}. Within the limits of measurement, no gas phase $^{13}$CO$_2$ was observed during the irradiation, indicating that the $^{13}$CO$_2$ remained on the surface. No additional $^{13}$CO$_2$ was produced during the TPDs because control experiments, where O$_3$ was irradiated with H before deposition of $^{13}$CO, yielded negligible $^{13}$CO$_2$, suggesting that when OH is produced, it reacts quickly on the surface.

When searched for, we saw no evidence of the production of H$^{13}$COOH, validating the hypothesis that investigating a submonolayer coverage restricts the chemistry to CO$_2$ production, unlike previous studies \citep{oba10, iop10}. Neither were H$_2^{13}$CO nor $^{13}$CH$_3$OH seen in desorption. However, H$_2$, $^{12}$CO, $^{13}$CO, $^{12}$CO$_2$ were seen during all TPDs; $^{13}$CO$_2$ was seen upon H irradiation; H$_2$O$_2$, H$_2$O were seen upon H irradiation, during extended TPDs to higher temperature; O$_3$ was seen only when O$_3$ had been deposited; O$_2$ was seen in all experiments, except those with O$_3$ and no H irradiation.

\begin{figure*}
\plotone{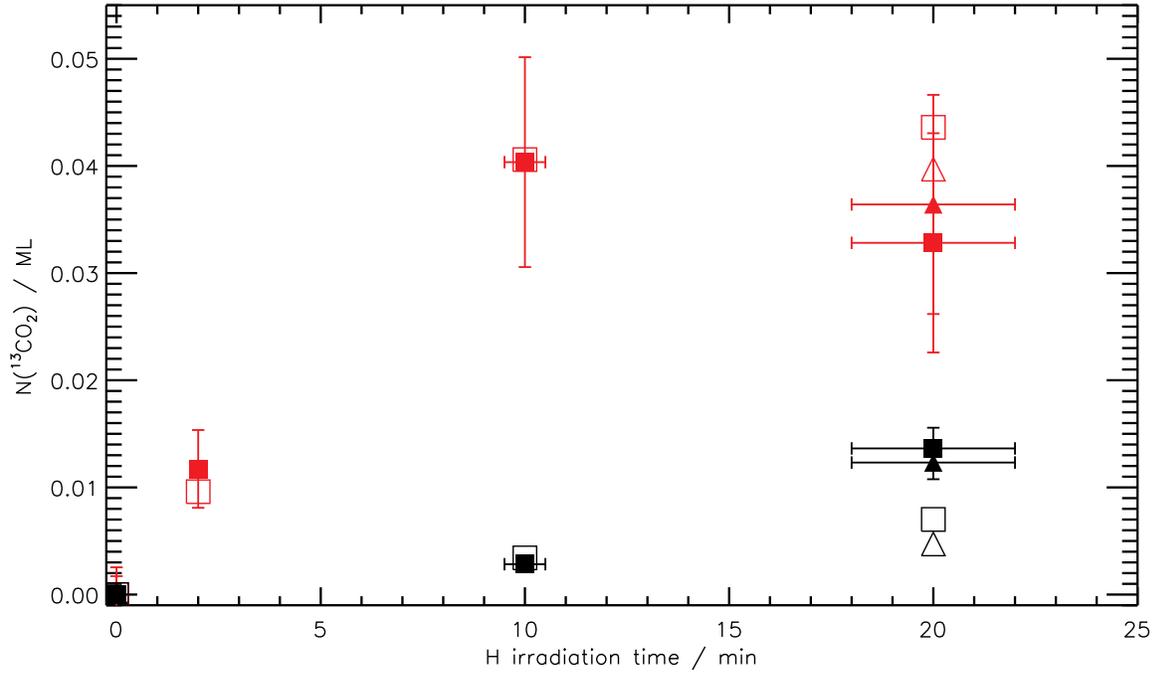}
\caption{The evolution of $^{13}CO_{2}$ with H irradiation time, presented in monolayers of CO$_2$, for all experiments. The data is labelled as follows: red, closed squares O$_3$ + $^{13}CO$ on \emph{np}-H$_2$O; red, closed triangles O$_3$ + $^{13}CO$ on silicate; black, closed squares O$_2$ + $^{13}CO$ on \emph{np}-H$_2$O; black, closed triangles O$_2$ + $^{13}CO$ on silicate. Overplotted (as corresponding open shapes) are the results of the kinetic model developed to describe the formation of $^{13}CO_{2}$. See text for details.\label{fig-co2}}
\end{figure*}

Figure~\ref{fig-co2} illustrates $^{13}$CO$_2$ production as a function of H irradiation time (solid symbols). It is clear that $^{13}$CO$_2$ was produced at comparable rates on both surfaces (solid triangles versus solid squares), regardless of the starting material (O$_2$ or O$_3$), vindicating our earlier conclusion that the underlying surface does not play a significant role in this reaction. In every experiment, some $^{13}$CO desorbed during the TPD, suggesting it was present in excess, never completely reacting with OH or H. In addition, as O$_3$, O$_2$ and H were dosed via one beam, while $^{13}$CO was dosed via a second, even after alignment the maximum overlap attainable is less than 100~\%, so not all reagents were dosed on the same region of the surface. Conversely, as Figure~\ref{fig-co2} shows, the reactions are very sensitive to the quantity of OH generated, which itself depends upon the starting quantity of O$_2$ or O$_3$ on the surface. Although surface coverages were controlled to within $\pm$~0.2~ML between experiments, this difference was sufficient to account for the varying concentration of $^{13}$CO$_2$ observed in Figure~\ref{fig-co2}.  

The maximum yield of $^{13}$CO$_2$ was $\sim$~8~\% with respect to $^{13}$CO (Table~\ref{tbl1}, experiments F,K); the presence of a complex barrier helps to explain this. Gas phase and theoretical studies predict a three-stage barrier ($\sim$~500~K, e.g. \citet{fro93,yu01}). The reaction proceeds via an energetic HOCO intermediate, which isomerises from \emph{trans}-HOCO to \emph{cis}-HOCO, before dissociating to form CO$_2$ \citep{smi73,ala93,les00}. \citet{les01} suggest that a precursor OH-CO complex forms prior to the HOCO intermediate. On a surface, the reaction probability is even more reliant upon the relative orientation of CO and OH. A recent theoretical study on a coronene surface shows that OH physisorbs with the hydrogen atom pointing towards the surface \citep{gou08}. Compared to the gas phase, the activation barrier to \emph{trans}-HOCO formation is slightly lowered, and the intermediate stabilised. A barrierless reaction between this stabilised HOCO complex with a further H atom could produce CO$_2$ + H$_2$. However, experiments suggest this reaction could also yield HCOOH or H$_2$O + CO \citep{iop11b}, while reaction with OH could yield H$_2$CO$_3$ \citep{oba10}. From the results presented here, it is not possible to determine whether $^{13}$CO$_2$ formed from \emph{cis}-HOCO $\rightarrow$ CO$_2$ + H, or by hydrogenation of HOCO, although no H$^{13}$COOH was observed when TPDs were run to 200~K, thus we assume that, under our experimental conditions, reaction (\ref{eqn:CO+OH}) produces only $^{13}$CO$_2$.

This complex barrier somewhat explains the low $^{13}$CO$_2$ yields, but there are further constraints to be considered. Due to the low coverages investigated here, the probability of $^{13}$CO and OH meeting on the surface is small. OH recombination could produce H$_2$O$_2$, or the competitive reaction OH + H could remove OH from the surface \citep{iop08}. Also, in these experiments, $^{13}$CO was dosed after O$_3$ or O$_2$, so at high enough coverages it could block H from reaching these reagents, allowing CO hydrogenation to artificially dominate. Previous studies show that hydrogenation of O$_3$ \citep{mok09} and O$_2$ \citep{iop08,dul10} occurs with no barrier, while hydrogenation of CO proceeds via:
\begin{equation}\label{eqn:CO+H}
  CO \xrightarrow{H}{} HCO \xrightarrow{H}{} H_2CO \xrightarrow{H}{} H_3CO \xrightarrow{H}{} CH_3OH
\end{equation} 
with a barrier of 390~K at 12~K to CO + H \citep{fuc09, wat02}. Here, the $^{13}$CO surface coverage was always below 1~ML, and neither H$_2^{13}$CO nor $^{13}$CH$_3$OH desorbed during extended TPDs, indicating that little $^{13}$CO hydrogenation occurred. If H$^{13}$CO was not produced in significant concentrations via (\ref{eqn:CO+H}), it follows that $^{13}$CO$_2$ was not produced at measurable quantities via (\ref{eqn:HCO+OH}), not least due to the low probability of any H$^{13}$CO produced encountering OH on the surface. These conclusions indicate that in this experiment $^{13}$CO$_2$ formation occurred exclusively via reaction (\ref{eqn:CO+OH}). 

\begin{figure*}
\includegraphics[width=0.45\linewidth]{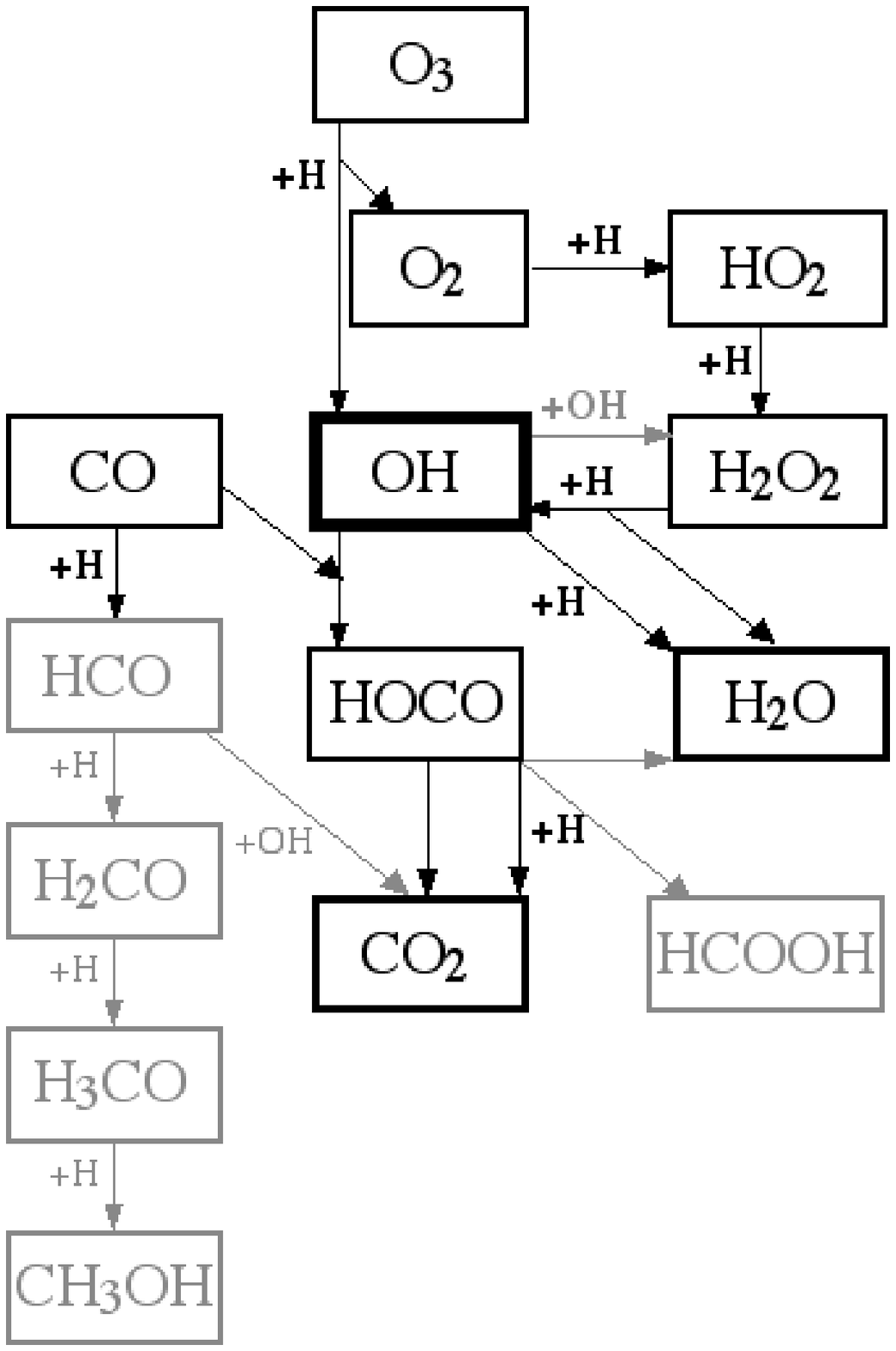}
\caption{Chemical network of OH as a schematic diagram. Reactions which occurred in the present study are depicted in black, whilst those that didn't are shaded grey. See text for details.\label{fig-network}}
\end{figure*}

A simple kinetic model of the experimental system was developed to describe the production of $^{13}$CO$_2$ via (\ref{eqn:CO+OH}), based on a set of coupled first order rate equations. Figure~\ref{fig-network} shows the potential detailed reaction scheme for these experiments. However, as illustrated above, given the low surface coverages employed here, a number of reactions, for example OH recombination, can be eliminated and are thus shown in grey in the figure. Neither CO$_2$ nor HCOOH react further with H \citep{bis07}, so such reactions were also ignored in the model. As discussed above, H$^{13}$CO was likely not produced at significant concentrations in these experiments. However, the rate of CO hydrogenation under present conditions was constrained by control experiments with only CO on the surface. It was assumed that the underlying surface had no effect on the reaction rate (as illustrated by Figure~\ref{fig-co2}), so the model treats both surfaces simultaneously. The best fit to the experimental data was found by varying the rates of reaction, k$_i$, of CO + OH and H$_2$O$_2$ + H, while constraining all other k$_i$ with empirical data (reactions shown in black in Figure~\ref{fig-network}). That only two free parameters (k$_{CO+OH}$ and k$_{H_2O_2+H}$) were required to fit all of the data presented here, provides strong evidence for the validity of the model, and our previous deductions that all $^{13}$CO$_2$ in our experiments was produced via reaction (\ref{eqn:CO+OH}). 

The results of the model are plotted over the experimental data as open symbols in Figure~\ref{fig-co2}, and are summarised in Table~\ref{tbl2}. At low H irradiation times, the model describes well the production of $^{13}$CO$_2$, but starts to deviate slightly at longer times. This could be attributed to route (\ref{eqn:HCO+OH}) becoming competitive, but in that case, the O$_3$ and O$_2$ experiments should deviate with the same trend, while in reality the production of $^{13}$CO$_2$ from O$_3$ is slightly overestimated while that from O$_2$ is underestimated. Reaction~(\ref{eqn:O3+H}) is exothermic \citep{mck55}, so the excess energy of this reaction could be transferred to the products, which may then desorb from the surface. The model does not account for this possibility, and thus overestimates the $^{13}$CO$_2$ yield by the O$_3$ route.

The model, while a simplistic approach to explaining the surface chemistry, as it ignores the complexity of the barrier known to exist in the CO + OH pathway, suggests that the relatively low yield of $^{13}$CO$_2$ in these experiments results from competition between (\ref{eqn:CO+OH}) and other reactions involving OH. The relation between key reaction rates is:
\begin{equation}\label{eqn:rate3}
  k_{CO+H} \ll k_{CO+OH} < k_{H_2O_2+H} < k_{OH+H},k_{O_2+H},
\end{equation}
indicating that water formation should always dominate the formation of ice species at low surface coverages. The overall effective reaction rate of CO + OH was determined to be 24 times slower than hydrogenation of OH, O$_3$ or O$_2$, and 1.7 times faster than CO + H. The relative rate between the hydrogenation of O$_2$ and CO was 40, while between the hydrogenation of O$_2$ and H$_2$O$_2$, it was 8, consistent with literature values of 31 -- 90 and 3.3, respectively \citep{miy08}. The factor of two difference between this and the previous value of $k_{O_2+H}/k_{H_2O_2+H}$ can be explained by the fact that molecular species such as H$_2$O$_2$ were more accessible to H in our (approximately 1~ML) experiments and therefore reacted more quickly than in the multilayer regime of \citet{miy08}.

It is not possible to extract the activation barrier for HOCO formation or its subsequent reactions to form CO$_2$ in reaction (\ref{eqn:CO+OH}) from the model. We can, however, calculate an effective barrier to CO$_2$ production by comparison of our modelled relative rates to the activation barrier of 390~K at 12~K for CO + H \citep{fuc09}. If we assume first order kinetics and a constant pre-exponential factor in both reactions, the effective barrier to (\ref{eqn:CO+OH}) is 384~$\pm$~40~K, where the error is derived from the barrier to hydrogenation of CO. This is the first calculation of the effective barrier to (\ref{eqn:CO+OH}) in the solid phase, and our result is contradictory to that of Oba et al (2010) who conclude that, in their study, the reaction proceeds with little or no activation barrier, but do not calculate a value. As suggested above, the presence of some fraction of excited OH in the beam could produce CO$_2$ in a barrierless reaction with CO. Current grain models include grain surface activation barriers to (\ref{eqn:CO+OH}) of e.g. 80~K (Y. Aikawa 2010, private communication) or 176~K \citep{cup09}, which would seem very low for an effective first order rate equation, potentially overproducing CO$_2$. All evidence suggests that CO + OH is a more efficient route to CO$_2$ formation than the non-energetic CO + O route, reaction (\ref{eqn:CO+O}), yet \citet{ros01} estimate the barrier to (\ref{eqn:CO+O}) is only 290~K, and suggest that the reaction proceeds under quiescent cloud conditions. If this were correct, then in comparison to the results of \citet{fuc09} and those presented here, CO + O would be more likely to proceed than either CO + H or CO + OH.

We feel, therefore, that it is relevant to address briefly the value of the CO + O barrier as derived by \citet{ros01} which, if implemented in gas-grain models, would result in the incorrect pathway to CO$_2$ formation dominating the reaction scheme. \citet{ros01} derived the barrier from an experiment where a water ice cap was deposited on top of CO which had previously been exposed to O atoms, assuming that reaction (\ref{eqn:CO+O}) occurred in the water pores as the surface was heated, and explicitly relying upon the reagents being trapped at the surface by the water ice cap. Although other laboratory studies show that CO trapping can occur in water ice pores \citep{col03}, observations and models confirm that CO freeze-out occurs after the formation of water ice layers in both molecular clouds \citep{pon03} and protostellar disks \citep{vis09}, implying that a water ice cap is not a realistic mimic of any interstellar ice, including those found in quiescent regions. In fact, scenarios investigating reaction (\ref{eqn:CO+O}) by \citet{ros01} under conditions comparable to those present in quiescent molecular clouds yielded no CO$_2$, nor did subsequent experimental studies by \citet{oba10}. Together with the effective barrier to (\ref{eqn:CO+OH}) of 384~$\pm$~40~K presented here, this suggests the importance of readdressing the value of the CO + O barrier implemented in astrochemical models.


\section{Astrophysical Implications}\label{sec-imp} 

The reaction of CO + OH is seen to be viable under astrophysical conditions of temperature and pressure, on silicate and \emph{np}-H$_2$O surfaces. Small quantities of CO$_2$ were produced, in competition with other reactions involving OH (for example hydrogenation of OH to form H$_2$O). Thus, the mechanism CO + OH could be key to explaining the formation of CO$_{2(s)}$ at the edges of dark clouds (low A$_V$), where CO$_{2(s)}$ is seen to form concurrently with H$_2$O on bare dust grains \citep{pon06}, before a large quantity of CO ice is present.

By modelling the reaction we have determined the empirical relationship $k_{CO+H} \ll k_{CO+OH} < k_{H_2O_2+H} < k_{OH+H},k_{O_2+H}$, where the overall effective rate of CO + OH is determined to be 24 times slower than OH + H, and 1.7 times faster than CO + H, indicating why H$_2$O ice is always the most abundant species.

In dense molecular clouds, gas phase hydrogen is observed to be mainly in the molecular form, but atomic H is present at a constant, low abundance (H/H$_2$~$\sim$~10$^{−3}$, \citet{li03}) due to the balance of H$_2$ formation on grain surfaces and its destruction by cosmic rays. The abundance of OH increases with density, in line with that of O (see e.g. \citet{har00,qua08}). Thus, after the freeze-out of CO, the reaction CO + OH could proceed on the ice mantle due to higher abundances of CO on the grain surface. As the abundance of OH increases, so does the potential for formation of CO$_2$ via CO + OH. The formation of H$_2$O via the competitive reaction OH + H will also increase with density, and thus CO$_2$ and H$_2$O formation in central, more quiescent regions of molecular clouds is possible. This conclusion agrees well with the postulations of \citet{gou08}.

\acknowledgments

We are grateful to Marius Lewerenz and David Field for helpful discussions. J.A.N. thanks The Leverhulme Trust, the Scottish International Education Trust, the University of Strathclyde and the Scottish Universities Physics Alliance for funding. The research leading to these results has received funding from the European Community's Seventh Framework Programme FP7/2007-2013 under grant agreement no. 238258. We acknowledge the support of the national PCMI programme founded by the CNRS, the Conseil Regional d'Ile de France through SESAME programmes (contract I-07-597R), the Conseil G\'{e}n\'{e}ral du Val d'Oise and the Agence Nationale de Recherche (contract ANR 07-BLAN-0129). We thank Jean-Louis Lemaire, Saoud Baouche and Henda Chaabouni.

\clearpage

\clearpage

\clearpage

\clearpage

\begin{deluxetable}{cccccc}
\tabletypesize{\scriptsize}
\tablecaption{Experiments performed in this work.\label{tbl1}}
\tablewidth{0pt}
\tablehead{
\colhead{Experiment} & \colhead{Substrate} & \colhead{N(O$_2$)} & \colhead{N(O$_3$)} & \colhead{N($^{13}$CO)} & \colhead{t(H)} \\ 
\colhead{(label)} & \colhead{} & \colhead{(ML)} & \colhead{(ML)} & \colhead{(ML)} & \colhead{(minutes)} \\ 
}
\startdata
A & H$_2$O & 0.5 & \nodata & 0.6$^\dag$ & 0 \\ 
B & H$_2$O & 0.5 & \nodata & 0.6$^\dag$ & 10 \\ 
C & H$_2$O & 0.5 & \nodata & 0.6$^\dag$ & 20 \\ 
\hline
D & H$_2$O & \nodata & 1.6 & 0.5 & 0 \\ 
E & H$_2$O & \nodata & 1.4 & 0.5 & 2 \\ 
F & H$_2$O & \nodata & 1.6 & 0.5 & 10 \\ 
G & H$_2$O & \nodata & 1.1 & 0.6 & 20 \\ 
\hline
H & Silicate & 0.45 & \nodata & 0.45 & 0 \\ 
I & Silicate & 0.45 & \nodata & 0.45$^\dag$ & 20 \\ 
\hline
J & Silicate & \nodata & 1.5 & 0.13 & 0 \\ 
K & Silicate & \nodata & 1.3 & 0.45 & 20 \\ 
\enddata
\tablecomments{Species were deposited on the surface in order from left to right, apart from those marked \dag, where $^{13}$CO was deposited first. All species except $^{13}$CO were deposited using the same beam (see \S\ref{sec-expt} for details.)}
\end{deluxetable}

\clearpage

\begin{deluxetable}{cc}
\tabletypesize{\scriptsize}
\tablecaption{Modelled relative rate constants.\label{tbl2}}
\tablewidth{0pt}
\tablehead{
\colhead{Reaction} & \colhead{k$_i$/k$_{O_2+H}$} \\ 
}
\startdata
O$_3$ + H & 1 \\ 
O$_2$ + H & 1 \\ 
HO$_2$ + H & 1 \\ 
H$_2$O$_2$ + H & 0.125$^\dag$ \\ 
CO + H & 0.025 \\ 
OH + H & 1 \\
CO + OH & 0.042$^\dag$ \\ 
\enddata
\tablecomments{\dag Rate was a free parameter in the model. All other rates were fixed, based on published empirical values detailed in the text.}
.\end{deluxetable}


\begin{thebibliography}{}

\bibitem[Alagia et al. (1993)]{ala93} Alagia, M., Balucani, N., Casavecchia, P., Stranges, D. \& Volpi, G.~G. \ 1993, \jcp, 98, 8341 
\bibitem[Amiaud et al. (2006)]{Amiaud06} Amiaud, L., Fillion, J.~H., Baouche, S., Dulieu, F., Momeni, A. \& Lemaire, J.~L. \ 2006, \jcp, 124, 094702 
\bibitem[Amiaud et al. (2007)]{ami07} Amiaud, L., Dulieu, F., Fillion, J.-H., Momeni, A., \& Lemaire, J.~L. \ 2007, \jcp, 127, 709
\bibitem[Baulch et al. (2005)]{bau05} Baulch, D.~L. et al. \ 2005, J Phys Chem Ref Data, 34, 757 
\bibitem[Bisschop et al.(2007)]{bis07} Bisschop, S.~E., Fuchs, G.~W., van Dishoeck, E.~F., \& Linnartz, H.\ 2007, \aap, 474, 1061
\bibitem[Chen \& Marcus (2005)]{chen05} Chen, W.~C. \& Marcus, R.~A. \ 2005, \jcp, 123, 094307 
\bibitem[Collings et al.(2003)]{col03} Collings, M.~P., Dever, J.~W., Fraser, H.~J., McCoustra, M.~R.~S., \& Williams, D.~A.\ 2003, \apj, 583, 1058 
\bibitem[Cuppen et al.(2009)]{cup09} Cuppen, H.~M., van Dishoeck, E.~F., Herbst, E., \& Tielens, A.~G.~G.~M.\ 2009, \aap, 508, 275
\bibitem[van Dishoeck et al. (1996)]{vanD96} van Dishoeck, E.~F., et al.\ 1996, \aap, 315, L349
\bibitem[Dulieu et al.(2010)]{dul10} Dulieu, F., Amiaud, L., Congiu, E., Fillion, J.-H., Matar, E., Momeni, A., Pirronello, V., \& Lemaire, J.~L.\ 2010, \aap, 512, A30
\bibitem[Ehrenfreund et al. (1997)]{ehr97} Ehrenfreund, P., Boogert, A.~C.~A., Gerakines, P.~A., Tielens, A.~G.~G.~M., \& van Dishoeck, E.~F.\ 1997, \aap, 328, 649
\bibitem[Frost, Sharkey \& Smith (1993)]{fro93} Frost, M.~J., Sharkey, P. \& Smith, I.~W.~M. \ 1993, \jcp, 97, 12254 
\bibitem[Fuchs et al. (2009)]{fuc09} Fuchs, G.~W., Cuppen, H.~M., Ioppolo, S., Romanzin, C., Bisschop, S.~E., Andersson, S., van Dishoeck, E.~F., \& Linnartz, H.\ 2009, \aap, 505, 629
\bibitem[Fujii et al. (2002)]{fuj02} Fujii, T., Selvin, P.~C., \& Iwase, K. \ 2002, Chem. Phys. Lett., 360, 367
\bibitem[Fulle et al. (1996)]{ful96} Fulle, D., Hamann, H.~F., Hippler, H. \& Troe, J. \ 1996, \jcp, 105, 983 
\bibitem[Gerakines et al. (1996)]{ger96} Gerakines, P.~A., Schutte, W.~A., \& Ehrenfreund, P.\ 1996, \aap, 312, 289
\bibitem[Gerakines et al. (1999)]{ger99} Gerakines, P.~A., et al.\ 1999, \apj, 522, 357
\bibitem[Gibb et al. (2004)]{gib04} Gibb, E.~L., Whittet, D.~C.~B., Boogert, A.~C.~A. \& Tielens, A.~G.~G.~M.\ 2004, \apjs, 151, 35  
\bibitem[Goumans et al. (2008)]{gou08} Goumans, T.~P.~M., Uppal, M.~A., \& Brown, W.~A.\ 2008, \mnras, 384, 1158
\bibitem[de Graauw et al. (1996)]{deG96} de Graauw, T., et al.\ 1996, \aap, 315, L345 
\bibitem[Greenberg (2002)]{gre02} Greenberg, J.~M.\ 2002, Surface Science, 500, 793
\bibitem[Harju et al.(2000)]{har00} Harju, J., Winnberg, A., \& Wouterloot, J.~G.~A.\ 2000, \aap, 353, 1065
\bibitem[D'Hendecourt \& Jourdain de Muizon (1989)]{dHendecourt89} D'Hendecourt, L.~B., \& Jourdain de Muizon, M.\ 1989, \aap, 223, L5
\bibitem[Ioppolo et al. (2008)]{iop08} Ioppolo, S., Cuppen, H.~M., Romanzin, C., van Dishoeck, E.~F., \& Linnartz, H.\ 2008, \apj, 686, 1474 
\bibitem[Ioppolo et al. (2009)]{iop09} Ioppolo, S., Palumbo, M.~E., Baratta, G.~A., \& Mennella, V.\ 2009, \aap, 493, 1017
\bibitem[Ioppolo et al.(2011a)]{iop10} Ioppolo, S., van Boheemen, Y., Cuppen, H.~M., van Dishoeck, E.~F., \& Linnartz, H.\ 2011, \mnras, Advance Online Publication. doi:10.1111/j.1365-2966.2011.18306.x
\bibitem[Ioppolo et al.(2011b)]{iop11b} Ioppolo, S., Cuppen, H.~M., van Dishoeck, E.~F., \& Linnartz, H.\ 2011, \mnras, 410, 1089 
\bibitem[Jamieson et al. (2006)]{jam06} Jamieson, C.~S., Mebel, A.~M., \& Kaiser, R.~I.\ 2006, \apjs, 163, 184 
\bibitem[Knez et al. (2005)]{kne05} Knez, C., et al.\ 2005, \apjl, 635, L145 
\bibitem[Laffon et al. (2010)]{laf10} Laffon, C., Lasne, J., Bournel, F., Schulte, K., Lacombe, S., \& Parent, Ph. \ 2010, PCCP, 12, 10865
\bibitem[Lemaire et al.(2010)]{lem10} Lemaire, J.~L., Vidali, G., Baouche, S., Chehrouri, M., Chaabouni, H., \& Mokrane, H.\ 2010, \apjl, 725, L156 
\bibitem[Lester et al. (2000)]{les00} Lester, M.~I., Pond, B.~V., Anderson, D.~T., Harding, L.~B. \& Wagner, A.~F., \ 2000, \jcp, 113, 9889
\bibitem [Lester et al. (2001)]{les01} Lester, M.~I., Pond, B.~V., Marshall, M.~D., Anderson, D.~T., Harding, L.~B., \& Wagner, A.~F. \ 2001, Faraday Discuss., 118, 373
\bibitem[Li \& Goldsmith(2003)]{li03} Li, D., \& Goldsmith, P.~F.\ 2003, \apj, 585, 823
\bibitem[McKinley et al.(1955)]{mck55} McKinley, J.~D., Garvin, D., \& Boudart, M.~J. \ 1955, \jcp, 23, 784
\bibitem[M{\'e}dard et al.(2002)]{med02} M{\'e}dard, N., Soutif, J.-C., \& Poncin-Epaillard, F. \ 2002, Langmuir, 18, 2246
\bibitem[Mennella et al.(2004)]{men04} Mennella, V., Palumbo, M.~E., \& Baratta, G.~A.\ 2004, \apj, 615, 1073 
\bibitem[Mennella et al.(2006)]{men06} Mennella, V., Baratta, G.~A., Palumbo, M.~E., \& Bergin, E.~A.\ 2006, \apj, 643, 923 
\bibitem[Miyauchi et al.(2008)]{miy08} Miyauchi, N., Hidaka, H., Chigai, T., Nagaoka, A., Watanabe, N., \& Kouchi, A.\ 2008, Chemical Physics Letters, 456, 27
\bibitem[Mokrane et al. (2009)]{mok09} Mokrane, H., Chaabouni, H., Accolla, M., Congiu, E., Dulieu, F., Chehrouri, M., \& Lemaire, J.~L.\ 2009, \apjl, 705, L195
\bibitem[Nummelin et al. (2001)]{num01} Nummelin, A., Whittet, D.~C.~B., Gibb, E.~L., Gerakines, P.~A., \& Chiar, J.~E.\ 2001, \apj, 558, 185 
\bibitem[Oba et al. (2010)]{oba10} Oba, Y., Watanabe, N., Kouchi, A., Hama, T., \& Pirronello, V.\ 2010, \apjl, 712, L174 
\bibitem[Oliveira et al. (2011)]{oli11} Oliveira, J.~M., et al.\ 2011, \mnras, 411, L36 
\bibitem[Palumbo et al. (1998)]{pal98} Palumbo, M.~E., Baratta, G.~A., Brucato, J.~R., Castorina, A.~C., Satorre, M.~A., \& Strazzulla, G.\ 1998, \aap, 334, 247 
\bibitem[Pontoppidan et al.(2003)]{pon03} Pontoppidan, K.~M., et al.\ 2003, \aap, 408, 981
\bibitem[Pontoppidan (2006)]{pon06} Pontoppidan, K.~M.\ 2006, \aap, 453, L47
\bibitem[Pontoppidan et al. (2008)]{pon08} Pontoppidan, K.~M., et al.\ 2008, \apj, 678, 1005
\bibitem[Quan et al.(2008)]{qua08} Quan, D., Herbst, E., Millar, T.~J., Hassel, G.~E., Lin, S.~Y., Guo, H., Honvault, P., \& Xie, D.\ 2008, \apj, 681, 1318
\bibitem[Roser et al. (2001)]{ros01} Roser, J.~E., Vidali, G., Manic{\`o}, G., \& Pirronello, V.\ 2001, \apjl, 555, L61
\bibitem[Ruffle \& Herbst(2001)]{ruf01} Ruffle, D.~P., \& Herbst, E.\ 2001, \mnras, 324, 1054
\bibitem[Shimonishi et al. (2010)]{shi10} Shimonishi, T., Onaka, T., Kato, D., Sakon, I., Ita, Y., Kawamura, A., \& Kaneda, H.\ 2010, \aap, 514, A12
\bibitem[Smith \& Zellner (1973)]{smi73} Smith, I.~W.~M. \& Zellner, R. \ 1973, J. Chem. Soc, Faraday Trans. II, 69, 1617 
\bibitem[Sun \& Law (2008)]{sun08} Sun, H. \& Law, C.~K. \ 2008, J. Mol. Struct.: THEOCHEM, 862, 138 
\bibitem[Talbi et al. (2006)]{tal06} Talbi, D., Chandler, G.~S., Rohl, A.~L. \ 2006, Chem. Phys., 320, 214
\bibitem[Visser et al.(2009)]{vis09} Visser, R., van Dishoeck, E.~F., Doty, S.~D., \& Dullemond, C.~P.\ 2009, \aap, 495, 881 
\bibitem[Watanabe \& Kouchi(2002)]{wat02} Watanabe, N., \& Kouchi, A.\ 2002, \apjl, 571, L173
\bibitem[Whittet et al.(1998)]{whi98} Whittet, D.~C.~B., et al.\ 1998, \apjl, 498, L159
\bibitem[Williams \& Herbst(2002)]{wil02} Williams, D., \& Herbst, E.\ 2002, Surface Science, 500, 823
\bibitem[Yu, Muckerman \& Sears (2001)]{yu01} Yu, H.~G., Muckerman, J.~T. \& Sears, T.~J. \ 2001, Chem. Phys. Lett., 349, 547 
\end{thebibliography}
\end{document}